\documentclass[12pt]{article}
\pdfoutput=1
\usepackage{graphicx}
\usepackage{amsmath}
\usepackage{amsfonts}
\usepackage{amssymb}
\usepackage{subcaption}
\usepackage{caption}
\usepackage[hidelinks]{hyperref}
\usepackage{mathrsfs}
\usepackage{enumerate}
\numberwithin{equation}{section}
\newcommand{\comment}[1]{}

\def\bsb{\boldsymbol}
\def\bea{\begin{eqnarray}}
\def\eea{\end{eqnarray}}
\def\be{\begin{equation}}
\def\ee{\end{equation}}

\comment{

\def\d{\partial}
\def\ep{\epsilon}

\def\P{\mathcal{P}}

\def\x{\bsb x}

\def\D{\mathcal{D}}
\def\vphi{\varphi}

}
\newcommand{\mal}[1]{\mathcal #1}
\newcommand{\expect}[1]{\left\langle #1 \right\rangle}

\def\H{\mal{H}}
\def\D{\mal{D}}
\def\O{\mathcal{O}}
\def\q{{\bsb q}}
\def\k{{\bsb k}}
\def\x{{\bsb x}}
\def\Re{\rm Re}

\def\d{\partial}
\def\ep{\epsilon}
\def\P{\mathscr{P}}
\def\zero{\bsb 0}
\def\qmax{q_{\rm max}}
\def\vphi{\varphi}
\begin{document}

\vspace*{0. cm}

\begin{center}

{\Large\bf Effective Theory of Squeezed Correlation Functions}

\vskip 1cm

{\large {\bf Mehrdad Mirbabayi} and {\bf Marko Simonovi\'c}}

\vskip 0.5cm

{\large {\em Institute for Advanced Study, 1 Einstein Drive, Princeton, NJ 08540}}

\end{center}

\vspace{.8cm}

\hrule \vspace{0.3cm}
{\small  \noindent \textbf{Abstract} \\[0.3cm]
\noindent Various inflationary scenarios can often be distinguished from one another by looking at the squeezed limit behavior of correlation functions. Therefore, it is useful to have a framework designed to study this limit in a more systematic and efficient way. We propose using an expansion in terms of weakly coupled super-horizon degrees of freedom, which is argued to generically exist in a near de Sitter space-time. The modes have a simple factorized form which leads to factorization of the squeezed-limit correlation functions with power-law behavior in $k_{\rm long}/k_{\rm short}$. This approach reproduces the known results in single-, quasi-single-, and multi-field inflationary models. However, it is applicable even if, unlike the above examples, the additional degrees of freedom are not weakly coupled at sub-horizon scales. Stronger results are derived in two-field (or sufficiently symmetric multi-field) inflationary models. We discuss the observability of the non-Gaussian 3-point function in the large-scale structure surveys, and argue that the squeezed limit behavior has a higher detectability chance than equilateral behavior when it scales as $(k_{\rm long}/k_{\rm short})^\Delta$ with $\Delta<1$ -- where local non-Gaussianity corresponds to $\Delta=0$. }

\comment{We propose an effective description of the squeezed limit correlation functions using an expansion in terms of a full basis of local operators. Through this expansion, the squeezed limit contains information about the active degrees of freedom during inflation and is constrained by unitarity. We derive those constraints. They distinguish Paolo inflation from galilean genesis. Stronger results can be obtained under certain assumptions. In particular, the degree to which the primordial power is the fluctuations of the inflaton in two-field (or sufficiently symmetric multi-field) models.
}
 \vspace{0.3cm}
\hrule

\begin{flushleft}
\end{flushleft}

\vspace{-1cm}
\section{Introduction}

One of the main goals of modern cosmology is to understand the origin of primordial fluctuations. We are fairly confident, given the in-phase acoustic oscillations, that the fluctuations have been super-horizon for a while. Inflation seems to be the simplest, and most widely accepted, explanation for generating the observed slightly red-tilted and almost Gaussian spectrum. However, there are many different inflationary scenarios, and there are alternatives to inflation, which we would like, and potentially be able to distinguish using data from observations in the foreseeable future. This program requires an understanding of the cosmological observables, and a classification of different models in terms of their distinctive observational signatures. 

Perhaps the most decisive of these signatures whose detection rules out non-inflationary alternatives is the $B$-type polarization of the cosmic microwave background caused by primordial tensor modes. However, depending on the energy scale of inflation, the amplitude of tensor modes may be too small for the $B$-modes to be observable, and even if observed, there is still much to learn about inflation. In particular, understanding what other degrees of freedom besides the inflaton are active during inflation and what is their spectrum of masses and couplings can teach us about the underlying ultraviolet physics. 

It is well known through several examples that the information about these degrees of freedom is encoded in the squeezed limit of correlation functions in momentum space, when one or a partial sum of momenta becomes ``soft'', i.e. much smaller in magnitude than the others. Single-field models of inflation satisfy a set of identities relating the squeezed limit correlation functions to lower order statistics \cite{Maldacena,Creminelli:2004yq}. In particular, denoting the soft momentum by $q$ and factoring out the power spectrum of the soft mode $P(q)$, the zeroth and first order terms in $q$ are fully fixed by these identities \cite{Creminelli_sct,Hinterbichler}. The model dependent self-interactions (for instance, the equilateral type 3-point correlators) start to contribute starting from second order. That is, neglecting the tensorial structure,
\be\label{equil}
\frac{\expect{\zeta_\q \cdots \zeta_\k}^{\rm nonuniversal}_{\rm single-field}}{P(q)}\propto q^2.
\ee
On the other hand additional light degrees of freedom during inflation leave characteristically different imprints in the squeezed limit correlation functions whose scaling in the ratio of the small momentum to the large ones $q/k$ is determined by the ratio of the mass to the expansion rate during inflation ($m/H$) \cite{Chen,Baumann}. The purpose of this note is to give a more unified treatment of this limit.

We take an approach similar in spirit to the effective field theory (EFT) of single-field inflation \cite{Cheung:2007st}. In the simplest single-field, slow-roll models of inflation the background expansion as well as the dynamics of scalar fluctuations are fully described in terms of one fundamental scalar field, the inflaton. However, there exist several models where the background is not driven by a weakly coupled fundamental field, such as DBI and $k$-inflation, but the fluctuations have a richer phenomenology (e.g. non-Gaussianity). These models are generally characterized by having a lower strong-coupling scale compared to the energy scale at which de Sitter isometries are broken \cite{Baumann_equil}. The advantage of the EFT of inflation is to provide a unifying framework to study the perturbations in these models by means of separating the discussion of background evolution from the fluctuations and parameterizing the former in a set of time-dependent coefficients. This allows a much more efficient and systematic treatment of the fluctuations, and in terms of parameters that are directly determined by observations. 

Similarly, the simplest inflationary models with more than one degree of freedom are usually described by the addition of new fields that are weakly coupled up to an energy scale $\Lambda \gg H$. They can be included in the EFT framework as in \cite{Baumann,Senatore_multi}, and their coupling to the inflaton, or their subsequent conversion to adiabatic modes modifies the phenomenology of inflation. However, distinct features generally appear in the behavior of the correlation functions in the squeezed limit. Away from this limit the shapes of correlation functions are expected to be degenerate with those resulting from self-interactions in a single-field model. Hence, although the EFT of multi-field inflation allows us to calculate the full shape of non-Gaussianity, only a part of it is of special interest. We therefore look for an effective description of this part by working directly at the level of the correlation functions rather than the Lagrangian. We will argue that such description exists, it is independent of details of interactions and is parameterized in terms of quantities that are directly related to observations. 

One can also imagine scenarios in which the additional degrees of freedom are strongly coupled below and around the Hubble scale, i.e. $\Lambda \sim H$. The EFT approach can fail in this case, while as we will argue, at super-horizon scales one seems to be able to always talk about weakly coupled degrees of freedom. This is a consequence of locality and the slow-roll assumption, namely that the time-variation of the parameters of the underlying theory is much slower than the expansion rate $H$. Therefore, one advantage of our approach is that it proves many of the known results about the squeezed limit behavior of inflationary correlation functions hold more universally, beyond the explicit models in which they were originally derived. In particular, the squeezed limit behavior is describable in terms of a discrete sum of power laws with either real positive or pairs of complex conjugate powers whose real parts are multiples of $3/2$. Any deviation (such as negative powers or a continuous spectrum) seems to indicate a radically different alternative to inflation.\footnote{In many of the existing alternatives to inflation the fluctuating fields see an effective de Sitter background metric (see \cite{Hinterbichler:2011qk} and references therein), and hence our conclusions remain unmodified.}

In what follows, we first investigate the evolution of local operators at super-horizon scales in more detail, and argue that they have simple scaling behavior as a function of the conformal time $\eta$, up to slow-roll corrections. In many cases of interest they become classical and their time and space dependence factorizes:
\be\label{O}
\sigma(\eta,x)\approx \hat \sigma(x) \eta^{\Delta_\sigma}.
\ee
We then show that to describe the squeezed limit behavior of the observed correlation functions, all that is needed is the spectrum of $\{\Delta_\sigma\}$, plus a set of coefficients parameterizing the degree of coupling between adiabatic fluctuations and each local operator, e.g.
\be
\begin{split}
\expect{\zeta(\x)\zeta(\zero)}_\sigma=&\xi_0(\x)+\hat\sigma(\zero) \xi_\sigma(\x)
+\d_i\hat\sigma(\zero)\xi^i_{\d\sigma}(\x)
+\d_i\d_j\hat\sigma(\zero)\xi^{ij}_{\d\d\sigma}(\x)+\cdots\\[10pt]
&+\hat\sigma^2(\zero)\xi_{\sigma^2}(\x)+\hat\sigma\d_i\hat\sigma(\zero)\xi^{i}_{\sigma\d\sigma}(\x)+\cdots
\end{split}
\ee
At this order the $\xi$ functions are fixed up to numerical coefficients which are the theoretical data directly related to the observations. This applies to quasi-single field and multi-field models but also when $\zeta$ is directly coupled to a composite operator like $\sigma^2$. The contribution of an operator of dimension $\Delta_O\geq 0$ to the squeezed limit scales as 
\be
\frac{\expect{\zeta_\q \cdots \zeta_\k}_O}{P(q)}\propto q^{\Delta_O}.
\ee
The special case in which the perturbations of the intermediate operator do not decay significantly by expansion ($\Delta_O \simeq 0$) deserves a more detailed discussion. These fluctuations are called entropic perturbations, which can convert to adiabatic perturbations at a later time during or after the end of inflation. The squeezed limit behavior is again simply related to a set of mixing and coupling coefficients. In the case where there is only one entropy mode, or there are several identical ones, we can determine the degree to which the observed adiabatic fluctuations are coming from the initial fluctuations of the inflaton, and derive a new consistency condition satisfied by ratios of squeezed limit correlation functions. In the more general case, unitarity imposes an inequality condition on these ratios \cite{Suyama,Smith}. 

Exceptions to \eqref{O} occur when there are massive particles of $m>3H/2$. This case is discussed separately since the corresponding super-horizon modes never become classical; they are described by two complex conjugate power laws and the notion of particle remains applicable. However, there is still a sense in which the correlation functions factorize in the squeezed limit which is analogous to that of scattering amplitudes on single particle poles. The squeezed limit of the bispectrum now scales as
\be
\frac{\expect{\zeta_\q \cdots \zeta_\k}_O}{P(q)}=\O(q^{3/2}),
\ee
but it also oscillates as a function of the ratio and the angle between the short and long momenta. This case is particularly interesting since the squeezed limit correlation functions, which in real space describe the correlation between short distance $\sim 1/k$ measurements performed at large separation $\sim 1/q$, have quantum mechanical origin in contrast to all previous cases. The oscillations arise from interference between two branches of the wavefunction \cite{Noumi,Arkani-Hamed}.

Finally, we discuss the observational prospects of these squeezed limit behaviors. It is known that local non-Gaussianity (a non-Gaussianity type that can be generated by entropy modes) has a higher chance of being detected by observation of ``non-local biasing'' in large-scale structure surveys \cite{Dalal}. This is because the squeezed limit behavior of local non-Gaussianity is distinct from what can be caused by gravitational dynamics in the formation of structures, which scales as $q^2$ in the soft momentum. In particular, by going to larger scales, or equivalently sending $q\to 0$ the contribution of local non-Gaussianity rises compared to the gravitational one. A milder enhancement can arise in all of the above scenarios. However, due to the finite size of the universe the amount of signal in the long wavelength modes is limited. Through an estimate of the signal-to-noise ratio we will argue that only those non-Gaussian 3-point functions that have the squeezed limit scaling below $\O(q)$ are better detectable in this limit.

\section{Super-horizon degrees of freedom}

Let us begin by considering an interacting theory of fields of non-zero mass during inflation. We denote them by $\{\sigma_a\}$, and order them with respect to their mass: $m_a\leq m_{a+1}$. As long as the time variation of coupling coefficients in the interaction Hamiltonian $\mathcal H$ is slower than the Hubble rate (the slow-roll assumption), it is easy to see that interactions become irrelevant for super-horizon modes. Hence, one can choose a basis of free fields. 

Note first that if one ignores the interactions among the sub-horizon and super-horizon modes and focus only on the latter, the terms with spatial derivative are negligible. Thus the equations of motion form a system of ODEs in time. Take the equation(s) for the lightest field(s). It is of the general form
\be\label{eom1}
a^{-3}\d_t(a^3 \dot\sigma_1)+m_1^2\sigma_1=F_1(\{\sigma_a\},t),
\ee
where $F_1$ is at least of second order in fields or their time-derivatives and it can explicitly depend on time through time-dependent coupling coefficients. Over-dot denotes $\d_t$. To argue that $F_1$ becomes negligible outside of the horizon we need to analyze the linear solutions. Up to slow-roll suppressed corrections, the super-horizon free-field solutions in a near de Sitter space-time of expansion rate $H$ are of the form
\be\label{sigmapm}
\sigma^\pm(\q,\eta)\propto \frac{(q\eta)^{\Delta_\sigma^{\pm}}}{q^{3/2}},\qquad 
\Delta_\sigma^{\pm}= \frac{3}{2}\pm\sqrt{\frac{9}{4}-\frac{m^2}{H^2}},
\ee
where $\eta = -1/aH$ is the conformal time and the $q$ dependence is fixed by the canonical quantization condition. It follows from the solution \eqref{sigmapm} that for masses below $3H/2$ the mode functions divide into growing and decaying modes. Therefore, after a few Hubble times the field fluctuations are dominated by a single growing mode and the spatial and temporal dependence of the perturbations factorize, giving
\be
\dot\sigma(\q,t)\simeq -\Delta_\sigma^{-} H \sigma(\q,t),
\ee
up to corrections that decay with a negative power of $a$. It follows that the field has become classical: by locality the equal-time real space commutator $[\sigma(\x,t),\dot\sigma(\x',t)]$ is proportional to $a^{-3}\delta^3(\x-\x')$ which implies that the momentum space commutator
\be\label{commut}
[\sigma({\q},t),\dot\sigma({\q'},t)]\propto a^{-3} \delta^3(\q+\q')
\ee
decays much faster than the product $\sigma({\q},t)\dot\sigma({\q'},t)\propto a^{-2\Delta_\sigma^-}$. Hence the field and its time derivative can be simultaneously known with negligible uncertainty. \comment{This super-horizon classicality of the light fields is expected to hold more generally as a consequence of locality, which ensures \eqref{commut}, and not to be limited to near de Sitter space-time.}For heavier fields with $m>3H/2$ this conclusion does not hold.

Returning to the main problem, the slow-roll assumption about the explicit time dependence of $F_1$ together with $m_a>0$ then imply that the interactions on the right-hand side of \eqref{eom1} decay by a negative power of $a$, and hence become negligible at late times. That is $\sigma_1$ becomes free. 

We can now proceed to the next field, say $\sigma_2$ with equation of motion given by
\be\label{eom2}
a^{-3}\d_t(a^3 \dot\sigma_2)+m_2^2\sigma_2=F_2(\{\sigma_a\},t).
\ee
Let us focus on one example, for instance, $F_2= \lambda \sigma_1^2$ (the generalization to arbitrary interaction is straightforward). If $\Delta_{\sigma_2}^- < 2\Delta_{\sigma_1}^-$, then the above argument can be repeated to show that $F_2$ becomes irrelevant at late times. However, in the opposite regime there can be interactions made of lighter fields which are not negligible. Here we can use the factorization property of the solutions for light fields to diagonalize the equation \eqref{eom2}. Note that $\Delta_{\sigma_2}^- \geq 2\Delta_{\sigma_1}^-$ can happen only if $\Delta_{\sigma_1}^- \leq 3/4$ for which the above-mentioned dominance of the growing mode and factorization of super-horizon solution is valid. We can therefore redefine the heavier field by adding to it a particular solution of the equation made of the lighter fields. In the above example, take
\be
\sigma_2 = \tilde\sigma_2 +\frac{\lambda}{2\Delta_1 (2\Delta_1 - 3)H^2+m_2^2}\sigma_1^2,
\ee
where $\Delta_1\equiv\Delta_{\sigma_1}^-$. The field $\tilde\sigma_2$ now satisfies a free field equation of mass $m_2$, plus corrections that are negligible outside of the horizon. This procedure can be continued to obtain a complete basis of free fields $\{\tilde \sigma_a\}$. The tildes will be dropped in the following.

So far the interactions with modes of the horizon size or shorter have been ignored. Once they are included and averaged over two types of corrections arise. First, their back-reaction to the long modes renormalizes the mass and coupling coefficients of the theory (or introduces new interactions), and secondly, the short modes can stochastically combine to make up a long wavelength mode. The first correction does not introduce any qualitative difference in the above argument except that the masses must be replaced by the renormalized ones. On the other hand, the stochastic contribution of horizon size modes on a super-horizon mode $q$ is suppressed because it requires $N\sim (aH/q)^3$ independent Hubble patches to conspire. The contribution to the variance of $\sigma$ is therefore expected to be of order $(q/aH)^3$ and hence negligible \cite{IR}.

Since every perturbative theory of massive fields approaches the free theory outside of the horizon, we expect any (possibly strongly coupled) theory which can be continuously connected to a perturbative theory by dialing its coupling coefficients to also satisfy this property. If there are massless fields that are not protected by a symmetry, the interactions do not necessarily decay with $a$. However, keeping these scalar fields massless seems to require considerable fine-tuning. We discard this possibility in what follows. Note that the metric fluctuations (both tensor and scalar modes) are massless, but their interactions contain sufficient number of derivatives and die off outside the horizon.  

It is important to stress that the above argument does not ensure that the super-horizon degrees of freedom continue to exist as weakly coupled fields at horizon and sub-horizon scales. However some general statements about correlation functions can be made even in these cases. For example, one can imagine that the physics at the sub-horizon scales is described by a strongly coupled conformal field theory (CFT).\footnote{One such example was proposed in \cite{cft}. There it was assumed that a sector of theory is described both at super- and sub-horizon scales by a CFT that is conformally coupled to the metric. We think one can consider a more general case where the coupling of the ultraviolet CFT to the metric is not conformal and hence it flows to a QFT with a mass gap of order $H$ on curved background. The above arguments would then apply in this case.} Or the compactification scale of the string theory may happen to be close to $H$ in which case a weakly coupled 4d description might not exist. The other interesting example is models with dissipation \cite{LopezNacir:2011kk}. In all these and similar scenarios it is hard or impossible to calculate the correlation functions. Nevertheless, the results of the following sections hold regardless of that. They are based just on the existence of free super-horizon degrees of freedom.

\section{Squeezed limit in a classical background field}

The result of the previous section has important consequences for the squeezed limit correlation functions. In order to see that let us consider a long-wavelength perturbation of a light field whose space and time dependence factorizes as argued above, $\sigma(\x,\eta)\simeq \hat\sigma(\x)g(\eta)$, where $g(\eta)$ is the growing solution. This long wavelength mode slightly modifies the background seen by the short wavelength fluctuations. Hence, in the squeezed limit of cosmological correlation functions, the contribution of $\sigma$ is through the response of shorter wavelength modes to this background. More specifically, this happens when the short modes start interacting with the long mode after its horizon crossing, or if their memory of earlier interactions is erased due to adiabatic evolution. In these cases an effective description can be employed. First, note that the short scale physics can only depend on local observables made of the long-wavelength mode which can be organized in a derivative expansion. For instance, the 2-point correlation functions of shorter wavelength metric fluctuations $\zeta$ on the background can be written as
\be\label{xi}
\begin{split}
\expect{\zeta(\x)\zeta(\zero)}_\sigma=&\xi_0(\x)+\hat\sigma(\zero) \xi_\sigma(\x)
+\d_i\hat\sigma(\zero)\xi^i_{\d\sigma}(\x)
+\d_i\d_j\hat\sigma(\zero)\xi^{ij}_{\d\d\sigma}(\x)+\cdots\\[10pt]
&+\hat\sigma^2(\zero)\xi_{\sigma^2}(\x)+\hat\sigma\d_i\hat\sigma(\zero)\xi^{i}_{\sigma\d\sigma}(\x)+\cdots
\end{split}
\ee
where $\xi_0$ is the correlation function in the absence of the particular long wavelength mode. This expansion in terms of local operators is not necessarily valid to all orders in the soft momentum $q$. For now, we are mainly interested in the leading contribution, however this point will be discussed in more detail in section \ref{sec:ope}. 

Equation \eqref{xi} implies that 
\be\label{z3xi}
\expect{\zeta_\q\zeta_{\k_1}\zeta_{\k_2}}=P_{\zeta\hat\sigma}(q)\xi_\sigma(\k_1)+\cdots
\ee
where $P_{\zeta\hat\sigma}(q)$ is the cross correlation $P_{\zeta\hat\sigma}(q)=\expect{\zeta_\q\hat\sigma_{-\q}}$ and dots represent other contributions to the squeezed limit 3-point function. If the mixing $r\equiv P_{\zeta\hat\sigma}/\sqrt{P_\zeta(q)P_{\hat\sigma}(q)}\ll 1$, then to leading order in $r$ we can write
\be\label{z3fac0}
\expect{\zeta_\q\zeta_{\k_1}\zeta_{\k_2}}_{\sigma}\simeq 
\frac{\expect{\zeta_\q\sigma_{-\q}(\eta_0)}\expect{\zeta_{\k_1}\zeta_{\k_2}\sigma_{\q}(\eta_0)}}
{P_{\sigma}(q,\eta_0)},
\ee
where $\eta_0$ is a time late enough so that all of the modes are super-horizon. The same formula \eqref{xi} can also be used to find the squeezed limit of a four-point function when the sum of two momenta is much less than individual momenta (this is sometimes called the counter-collinear limit). We get (for $\k_1\approx -\k_2$ and $\k_3\approx -\k_4$)
\be\label{z4fac}
\expect{\zeta_{\k_1}\zeta_{\k_2}\zeta_{\k_3}\zeta_{\k_4}}=\sum_{\vphi_1,\vphi_2\in\{\zeta,\hat\sigma_a\}}
P_{\vphi_1\vphi_2}(q)\xi_{\vphi_1}(\k_1)\xi_{\vphi_2}(\k_3)+\cdots
\ee

Note that the background long wavelength mode can be a fluctuation in $\zeta$ (or the tensor modes $\gamma_{ij}$). However, unlike the generic case, now there are consistency conditions on how the short modes respond to a long-wavelength metric fluctuation. Some linear superpositions of these fluctuations are locally indistinguishable from coordinate transformations and the response of the short modes is completely fixed by symmetries \cite{Hinterbichler}. Model dependence starts to show up at order $\d_i\d_j\zeta$ which is the reason for the squeezed limit behavior of equilateral-type shapes \eqref{equil}. In the absence of other super-horizon degrees of freedom the above constraints lead to consistency conditions on the full squeezed limit correlation functions, which hold regardless of the details of sub-horizon physics. For instance, an explicit derivation when the sub-horizon dynamics is effectively dissipative can be found in \cite{Nacir}.

In the presence of other degrees of freedom, two cases are of special interest and will be discussed separately in the following sections. First, when $\sigma$ has mass comparable to $H$ (or more precisely $m^2/H^2\gg 1-n_s$, the scalar tilt). In this case $\sigma$ fluctuations gradually decay outside the horizon, and the long-short correlation \eqref{xi} forms around the horizon crossing time of the short modes during inflation. The second case is when $m/H\ll 1$. In this case \eqref{xi} can form simultaneously for all $k$, long after they have all crossed the horizon. We denote these nearly massless fluctuations as entropy modes.

\subsection{Massive fields with $1-n_s\ll m^2/H^2 < 9/4$}\label{sec:quasi}

Neglecting slow-roll corrections, the growing mode of a field with mass $m<3H/2$ at super-horizon scales behaves as
\be\label{sigq}
\sigma_\q(\eta)\simeq\hat \sigma_\q \eta^{\Delta_\sigma^-}\propto \frac{(q\eta)^{\Delta_\sigma^{-}}}{q^{3/2}}.
\ee
Since the correlation \eqref{xi} forms around the horizon crossing time $\eta\sim -1/k$, the factor $\eta^{\Delta_\sigma^{-}}$ translates into $k^{-\Delta_\sigma^{-}}$. On the other hand, the scale invariance of de Sitter (i.e. the symmetry under $(\eta,\x)\to (\lambda\eta,\lambda\x)$) implies that
\be
\expect{\zeta_\q \sigma_{-\q}(\eta_0)}'\propto \frac{(q\eta_0)^{\Delta_\sigma^{-}}}{q^3}.
\ee
From \eqref{sigq} we see that $P_\sigma(q,\eta_0)\propto (q\eta_0)^{2\Delta_\sigma^{-}}/q^3$ up to normalization constants and $\expect{\sigma_\q(\eta_0)\hat\sigma_{-\q}}=P_\sigma(q,\eta_0)/\eta_0^{\Delta_\sigma^{-}}$. Hence, the $q$ dependence of \eqref{z3xi} and \eqref{z3fac0} are fully fixed and using the scale invariance we can write
\be\label{quasi}
\expect{\zeta_\q\zeta_{\k_1}\zeta_{\k_2}}_{\sigma}\propto P_\zeta(q)P_\zeta(k) \left(\frac{q}{k}\right)^{\Delta_\sigma^{-}},
\ee
where in this approximation $P_\zeta(q)\propto q^{-3}$. It is seen that a wide range of squeezed limit behaviors with $0<\Delta_\sigma^{-} <3/2$ result depending on the mass of $\sigma$. The quasi-single field scenario \cite{Chen:2009zp} is the perturbative realization of this possibility. A review of the perturbative calculation is provided in appendix \ref{funda}. 

Note that there can also be a mixing between $\zeta$ and a composite operator made of $\sigma$. For instance, if 
\be
P_{\zeta\hat\sigma^2}(q)\equiv\expect{\zeta_\q\hat\sigma^2(\q)}'\neq 0
\ee
then there will be a contribution like \eqref{quasi} but with $\Delta_\sigma^{-}\to 2\Delta_\sigma^{-}$. More generally for any local operator $O$ of dimension $\Delta_O\leq 2$, we have the squeezed limit behavior $(q/k)^{\Delta_O}$ that decays slower than the equilateral-type contributions. This is explicitly demonstrated in a perturbative example in appendix \ref{operator}. Note that the whole range between the local and equilateral type non-Gaussianity is covered in these models. This type of correlation between the metric fluctuation $\zeta$ and operators constructed out of super-horizon degrees of freedom can naturally arise, for instance via a coupling $\dot\zeta \sigma^2$.

\subsubsection{Squeezed limit and operator product expansion\label{sec:ope}}

Before discussing the case of entropic fluctuations, let us make some comments which cast doubt on any general connection between squeezed limit correlation functions during inflation (or a near de Sitter phase) and operator product expansion (OPE) in a hypothetical conformal field theory in one lower dimension (hints that such a connection might exist can be found e.g.~in \cite{Arkani-Hamed,Assassi}). The presence of such a connection is motivated by the fact that the action of de Sitter isometries on super-horizon fields can be identified with the action of conformal group on some dual operators in a CFT living on a spatial slice. This can best be seen at the level of the wavefunction of the universe. The wavefunction is most naturally expressed for fields of dimension lower than $3/2$ in terms of sufficiently smooth (super-horizon) field configurations as $\Psi[\{\vphi\},\eta]$. This is the analog of coordinate basis for particle quantum mechanics $\psi(x,t)$, but now with one real variable for each $\vphi(\x)$, or for each $\vphi(\q)$ with $q\ll -\eta^{-1}$. The expectation values are calculated by integrating over all of these fields 
\be
\expect{F[\{\vphi(\eta)\}]}=\int D\vphi |\Psi[\{\vphi\},\eta]|^2 F[\{\vphi\}].
\ee
Note that $\vphi$ on the r.h.s. is an integration variable and the time-dependence is fully encoded in $\Psi$. 

The coefficient of the term $\vphi_1\cdots \vphi_n$ in $\log \Psi$ have the same structure as an $n$-point correlation function of CFT operators $\expect{O_1\cdots O_n}$ with dimensions $\Delta_{O_i}=\Delta^{+}_{\vphi_i}=3-\Delta^{-}_{\vphi_i}$ times $\eta^{\Delta_1^{-}}\cdots \eta^{\Delta_n^{-}}$. (For scalar fields heavier than $3H/2$ the mapping is less clear since both asymptotic solutions, $\Delta^{\pm}_\vphi$, decay the same way.) Now one may expect that the dominant squeezed limit behavior of correlation functions should be related to the OPE limit of the CFT operators, where two operators dual to the short modes are brought close to each other in position space, and their product is expanded as a sum over operators. We will show that this is not necessarily the case.

\begin{figure}
\includegraphics[width=0.99\textwidth]{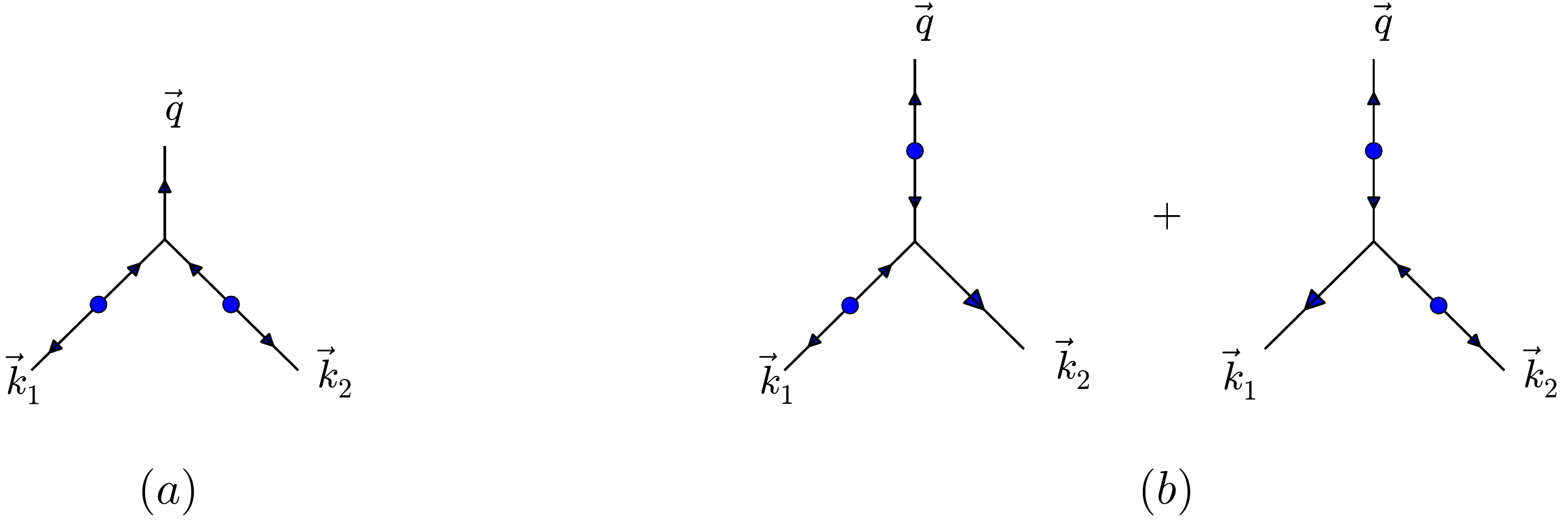}
\caption{(a) The diagram that corresponds to the squeezed limit bispectrum calculated using OPE (see equation \eqref{OPE}). Both power spectra are on the short modes. (b) Diagrams for the leading part of the squeezed limit bispectrum given by \eqref{phiphisigma}. One of the power spectra is on the long mode. }
\label{fig:diagrams}
\end{figure}

As a concrete example consider $\phi(\x_1)\phi(\x_2)\sigma(\x_3)$ where $\phi$ is a dimension zero field (an approximation to the almost massless fluctuations of the inflaton). The coefficient of this term has the structure of $\expect{\Phi(\x_1)\Phi(\x_2)\Sigma(\x_3)}$, where $\Delta_\Phi = 3$ and $\Delta_\Sigma = \Delta_\sigma^+$. In the limit $|\x_{12}|=|\x_1-\x_2|\to 0$, one expects
\be
O_1(\x_1)O_2(\x_2)\sim \sum_{Q} c_{12}^{\Delta_Q} x_{12}^{\Delta_Q-\Delta_1-\Delta_2} Q(\x_1),
\ee
where the constants $c_{12}^{\Delta_Q}$ are called OPE coefficients. Using $\expect{Q(\x_1)Q(\x_2)}\propto x_{12}^{-2\Delta_Q}$ gives a term in $\log \Psi$ of the form
\be
\int \prod_i d^3\x_i\phi(\x_1)\phi(\x_2)\sigma(\x_3) 
x_{12}^{\Delta^{+}_\sigma-6}x_{13}^{-2\Delta_\sigma^{+}}.
\ee
Taking the Fourier transform, we obtain
\be\label{f(k,q)}
\int d^3\k d^3\q \phi(\k)\phi(-\k-\q)\sigma(\q) f(k,q,\eta),\quad \text{with}\quad 
f(k,q,\eta)\propto k^{3-\Delta^{+}_\sigma} q^{2\Delta^+_\sigma-3} \eta^{-\Delta_\sigma^-}.
\ee
The 3-point correlation function can be computed via
\be
\expect{\phi_{\k_1}(\eta)\phi_{\k_2}(\eta)\sigma_{\k_3}(\eta)}=
\int D\phi D\sigma \phi(\k_1)\phi(\k_2)\sigma(\k_3)|\Psi[\{\vphi\},t]|^2.
\ee
For small non-Gaussianities, $|\Psi|^2$ is dominated by the Gaussian part
\be
|\Psi[\{\vphi\},\eta]|^2=\exp\left[\sum_{\{\vphi\}} -\frac{1}{2}\int d^3\k P_\vphi^{-1}(\k,\eta)\vphi(\k)\vphi(-\k)\right],
\ee
and the non-Gaussian correlation functions can be calculated perturbatively. The cubic term \eqref{f(k,q)} gives the squeezed limit contribution
\be\label{OPE}
\expect{\phi_{\k}(\eta)\phi_{-\k-\q}(\eta)\sigma_{\q}(\eta)}_{\rm OPE}
=f(k,q,\eta)P_\phi^2 (k)P_\sigma(q,\eta)\propto \eta^{\Delta_\sigma^-} k^{3-\Delta^{+}_\sigma}P_\phi^2(k),
\ee
where we have used the fact that 
\be
P_\sigma(q,\eta) \propto \frac{(q\eta)^{2\Delta_\sigma^-}}{q^3}= \eta^{2\Delta_\sigma^-} q^{2\Delta_\sigma^+ -3}.
\ee
However, one can show that \eqref{OPE} is a subleading contribution in the squeezed limit. The leading contribution comes from correlating with $\sigma_\q(\eta)$ a similar expression as \eqref{xi} with the substitution $\zeta\to \phi$. Up to an overall constant, this is fixed by the scale invariance 
\be\label{phiphisigma}
\expect{\phi_\k\phi_{-\k-\q}\sigma_\q(\eta)}
\propto \eta^{-\Delta_\sigma^-} k^{-\Delta^{-}_\sigma}P_\phi(k)P_\sigma(q,\eta).
\ee
(Note that super-horizon $\phi_\k$ is time-independent in our approximation.) Hence
\be\label{ope/squeezed}
\frac{\expect{\phi_\k\phi_{-\k-\q}\sigma_\q(\eta)}_{\rm OPE}}{\expect{\phi_\k\phi_{-\k-\q}\sigma_\q(\eta)}}
\sim \left(\frac{q}{k}\right)^{3-2\Delta_\sigma^{-}}\ll 1,
\ee
where we used the fact that $\Delta_\sigma^{-}<3/2$. Note that on the level of in-in diagrams the OPE result corresponds to having both power spectra on the short modes, while the diagrams for expression \eqref{phiphisigma} have one of the power spectra on the long mode. These different diagrams are shown in Fig.~\ref{fig:diagrams}.

The dominant squeezed limit contribution \eqref{phiphisigma} arises from the following cubic term in $\log \Psi$:
\be
\int d^3\k d^3\q \phi(\k)\phi(-\k-\q)\sigma(\q) 
k^{-\Delta^{-}_\sigma} P_\phi^{-1}(k) \eta^{-\Delta_\sigma^-}.
\ee
This is a local function of $x_{13}$ when transformed to the position space. Interpreted as a correlation function of dual operators, it is a contact term $\expect{\Phi(\x_1)\Phi(\x_2)\Sigma(\x_3)}_{\rm contact} \propto \delta^3(\x_1-\x_3)$. However, this does not make it any less relevant for cosmological correlation functions. We can translate to correlation functions of $\zeta$ by using
\be
\zeta \simeq -\frac{H\phi}{\dot\phi_{\rm s.r.}},
\ee
where $\dot\phi_{\rm s.r.}$ is the background slow-roll rate of the change of $\phi$. The field $\sigma$ would then contribute to the 3-point function if there is a mixing $P_{\zeta\sigma}$, and, even if not, its exchange contributes to the squeezed 4-point function ($|\q|=|\k_1+\k_2|\ll k_i$):
\be
\expect{\zeta_{\k_1}\zeta_{\k_2}\zeta_{\k_3}\zeta_{\k_4}}_\sigma=\left(\frac{H}{\dot\phi_{\rm s.r.}}\right)^4
\frac{\expect{\phi_{\k_1}\phi_{\k_2}\sigma_\q(\eta)}'\expect{\phi_{\k_3}\phi_{\k_4}\sigma_\q(\eta)}'}{P_\sigma(q,\eta)}.
\ee
The resulting three and four point functions are non-analytic functions of $q$ which lead to long-range correlations in position space. 

Our expansion \eqref{xi} keeps only these local terms in $\log\Psi$, which as argued above play the dominant role in the squeezed limit. The corrections generically start at order estimated in \eqref{ope/squeezed}.

\subsection{Entropic perturbations $m/H \ll 1$}

The fluctuations of a very light field $\sigma$ remain almost constant during inflation and the power in fluctuations is nearly scale invariant. As a consequence they can affect cosmological observables at a later epoch of cosmic evolution, e.g. during reheating or when the $\sigma$ fluctuations become time-dependent again (for instance when the expansion rate drops below their mass). In the latter case the mode functions are not given by \eqref{sigmapm} with $\Delta^-\simeq 0$ anymore, however we will argue that the modes remain classical and to study the affect of super-horizon perturbations of $\sigma$ on the observables one can still use an expansion in terms of local observables \eqref{xi}. 

Suppose at some time the solutions to the linear equation of motion for $\sigma$ become oscillatory with a (time-dependent) physical frequency $\omega$. Now both mode functions have to be kept but the reason why the field can still be treated classically is that the fluctuations have very large occupation number. The amplitude of the inflationary fluctuations are $\expect{\sigma_k^2}'\sim H_{\rm inf}^2/k^3$ while the typical size of the vacuum fluctuations for a fluctuating field of frequency $\omega$ is $\expect{\sigma_k^2}_{\rm vac}\sim 1/a^3\omega$. The ratio of the two is a measure of the occupation number which is 
\be
N \sim \frac{\expect{\sigma_k^2}}{\expect{\sigma_k^2}_{\rm vac}}\sim \frac{\omega H_{\rm inf}^2}{(k/a)^3}.
\ee
This is much larger than unity if the mode starts oscillating while it is still super-horizon, in which case $k/a\ll \omega\ll H_{\rm inf}$, and also when it enters the (sound) horizon long after the end of inflation, when $\omega = k/a\sim H_{\rm reentry} \ll H_{\rm inf}$. This implies that $\sigma$ and $\dot\sigma$ can both be known with great accuracy and the evolution of the field is determined in terms of its freeze-out value during inflation $\hat \sigma$. As long as the $\sigma$ mode is super-horizon this evolution is also local and therefore an expansion like \eqref{xi} can be used to relate observables to the locally measurable operators made of $\sigma$ which can all be related to the local value of $\hat \sigma$ and its derivatives. Moreover, since the interactions occur when all modes are already super-horizon they should be independent of $k$, therefore the squeezed limit behavior is the same as local non-Gaussianity. Note that, as before this local non-Gaussianity can be generated via a mixing between $\zeta$ and $\sigma^2$ or other composite operators made of $\sigma$---energy density is perhaps the most natural candidate---while $P_{\zeta\sigma}$ can in principle vanish.


These models are known to violate standard single-field consistency conditions since unlike long wavelength $\zeta$ perturbations those of $\sigma$ are not locally equivalent to a coordinate transformation. That is, while in the squeezed limit
\be\label{single_cons}
\xi_{\zeta_\q}(\k_1,\cdots,\k_N)\to \D\expect{\zeta_{\k_1}\cdots\zeta_{\k_N}},
\ee
where $\D$ is the dilatation operator, $\xi_{\sigma_\q}(\k_1,\cdots,\k_N)$ is completely model dependent. However, we can still find weaker conditions which are satisfied by models in which the violation of the single-field relations is triggered by mixing with other fields, thereby distinguishing them from models in which $\zeta$ by itself evolves at super-horizon scales.\footnote{The latter case can be characterized by the condition that the constant adiabatic mode of Weinberg \cite{Weinberg_adia} is a sub-dominant solution to the linearized field equations.}

The idea is to use the knowledge about the vertices with soft $\zeta$ modes [like \eqref{single_cons}], and the factorization to eliminate the ignorance about interactions of soft $\sigma$ modes. We first consider only a single entropic mode, which is easily generalizable to the case of several identical ones. Here, one can fully determine the degree of mixing between $\sigma$ and $\zeta$ from combinations of observable squeezed-limit correlation functions. The mixing cannot exceed $1$ by unitarity, otherwise the two-by-two correlation matrix will have negative eigenvalues. Moreover, different ways of determining it from different combinations of correlation functions must agree. This gives a consistency condition for these models. In the more general case, the unitarity requirement still puts a weaker constraint on the ratio of squeezed limit correlation functions. Most notably, on $f_{NL}$ and $\tau_{NL}$, which characterize respectively the 3-point and the 4-point correlators when the sum of two momenta becomes soft. However, these weaker constraints on observables must be obeyed as a consequence of the axioms of probability \cite{Smith}, and therefore any consistent theory of fluctuations would automatically satisfy them.

\subsubsection{Consistency conditions on two-field scenarios}

Consider having only one entropy mode. In this case, we have
\be
\expect{\zeta_\q\zeta_{\k_1}\zeta_{\k_2}}'=P_{\zeta\hat\sigma}(q)\xi_\sigma(\k_1)+P_{\zeta}(q)\xi_{\zeta}(\k_1),
\ee
and 
\be
\begin{split}
\expect{\zeta_{\k_1}\zeta_{\k_2}\zeta_{\k_3}\zeta_{\k_4}}'=&
\xi_{\zeta}(\k_1)\xi_{\zeta}(\k_3)P_\zeta(q)
+\xi_{\sigma}(\k_1)
\xi_{\sigma}(\k_3)P_{\hat\sigma}(q)\\[10pt]
&+[\xi_{\zeta}(\k_1)\xi_{\sigma}(\k_3)+\xi_{\sigma}(\k_1)\xi_{\zeta}(\k_3)]P_{\zeta\hat\sigma}(q),
\end{split}
\ee
for the counter-collinear limit of the 4-point function $\k_1+\k_2=\q\to0$. Solving for $r = P_{\zeta\hat\sigma}/\sqrt{P_\zeta(q)P_{\hat\sigma}(q)}$, we obtain\footnote{Notice that in this equation we treat $r$ non-perturbatively.}
\be
\label{eq:r2_nonpert}
r^2 =\frac{[\expect{\zeta_{\k_1}\zeta_{\k_2}\zeta_{\q}}'-P_{\zeta}(q)\xi_{\zeta}(\k_1)]
[\expect{\zeta_{\k_3}\zeta_{\k_4}\zeta_{-\q}}'-P_{\zeta}(q)\xi_{\zeta}(\k_3)]}
{P_\zeta(q)[\expect{\zeta_{\k_1}\zeta_{\k_2}\zeta_{\k_3}\zeta_{\k_4}}'
- \expect{\zeta_{\k_1}\zeta_{\k_2}\zeta_{\q}}' \xi_{\zeta}(\k_3)
-\expect{\zeta_{\k_3}\zeta_{\k_4}\zeta_{\q}}' \xi_{\zeta}(\k_1)
+P_\zeta(q) \xi_{\zeta}(\k_1)\xi_{\zeta}(\k_1)]}.
\ee
This must hold for all $\k_1$ and $\k_3$ as long as they are much larger than $\q$.\footnote{One may wonder if the denominator can be equal to zero for some momentum configurations. It is equal to $P_{\hat\sigma}(q) P_{\zeta}(q) \xi_{\sigma}({\bf k}_1) \xi_{\sigma}({\bf k}_3)$ and therefore it can vanish only when the numerator vanishes as well. Note that in this case $r$ cannot be determined using \eqref{eq:r2_nonpert}.} Using the standard consistency condition \eqref{single_cons}, and assuming that $r\ll 1$ simplifies the above equation to
\be\label{z3/z4}
r^2=\frac{[\expect{\zeta_{\k_1}\zeta_{\k_2}\zeta_{\q}}'-P_\zeta(q)\D P_\zeta(k_1)]
[\expect{\zeta_{\k_3}\zeta_{\k_4}\zeta_{-\q}}'-P_\zeta(q)\D P_\zeta(k_3)]}
{P_\zeta(q)\expect{\zeta_{\k_1}\zeta_{\k_2}\zeta_{\k_3}\zeta_{\k_4}}'
- P_\zeta^2(q)\D P_\zeta(k_1)\D P_\zeta(k_3)}.
\ee

The above equation tells us that, even though the $\sigma$ field itself is unobservable, the mixing parameter $r$ can still be measured just from the observable correlation functions of $\zeta$. Moreover, $r$ can also be solved in terms of the counter-collinear limit of the 5-point function and the squeezed limit of 3- and 4-point functions, and so on. Then, given that the two expressions for $r$ must agree, one can write down an equality that involves just the correlators of $\zeta$. These relations are new consistency conditions for the two-field inflation that hold irrespectively of the details of $\sigma$ interactions.

\subsubsection{Unitarity constraints on multi-field models}

In the more general case where a set of local operators $\{O_a\}$ contribute to the squeezed limits, the constraints are weaker, since in general one cannot solve for the individual mixings $r_a$. However, there are inequalities that must be satisfied for the theory to be unitary.\footnote{For a discussion of unitarity constraints on single field inflation see \cite{anal}.} We derive the simplest one: The squeezed limit of the 3-point function and the counter-collinear limit of 4-point function now read
\be
\expect{\zeta_\q\zeta_{\k_1}\zeta_{\k_2}}'=\sum_a P_{\zeta a}(q)\xi_a(\k_1)+P_{\zeta}(q)\xi_{\zeta}(\k_1),
\ee
and 
\be
\begin{split}
\expect{\zeta_{\k_1}\zeta_{\k_2}\zeta_{\k_3}\zeta_{\k_4}}'=&
\xi_{\zeta}(\k_1)\xi_{\zeta}(\k_3)P_\zeta(q)
+\sum_{a,b} \xi_{a}(\k_1)
\xi_{b}(\k_3)P_{ab}(q)\\[10pt]
&+\sum_a [\xi_{\zeta}(\k_1)\xi_{a}(\k_3)+\xi_{a}(\k_1)\xi_{\zeta}(\k_3)]P_{\zeta a}(q),
\end{split}
\ee
where $P_{ab}=\expect{\hat O_a\hat O_b}$. Choosing an orthogonal basis for the operators such that $P_{ab}\propto\delta_{ab}$, and assuming for simplicity that $\k_1\simeq \k_3$, we arrive at the following constraint (using the Cauchy-Schwartz theorem)
\be
[\expect{\zeta_{\k_1}\zeta_{\k_2}\zeta_{\q}}'-P_{\zeta}(q)\xi_{\zeta}(\k_1)]^2\leq
(\sum_a r_a^2)
P_\zeta(q)[\expect{\zeta_{\k_1}\zeta_{\k_2}\zeta_{\k_3}\zeta_{\k_4}}'
- 2\expect{\zeta_{\k_1}\zeta_{\k_2}\zeta_{\q}}' \xi_{\zeta}(\k_1)
+P_\zeta(q) \xi_{\zeta}(\k_1)\xi_{\zeta}(\k_1)].
\ee
In the extremely squeezed limit where only local shape survives, this simplifies to
\be\label{tildfNL}
\left(\frac{12}5 \tilde f_{NL}\right)^2\leq (\sum_a r_a^2)\left[4\tau_{NL}-\left(\left(\frac{12}5 f_{NL}\right)^2-\left(\frac{12}5 \tilde f_{NL}\right)^2\right) \right] \;,
\ee
where $\tilde f_{NL}$ is the observable part of the local non-Gaussianity: the difference between total $f_{NL}$ and the single-field part. A stronger bound can be obtained by subtracting the contribution of tensor mode exchange from the 4-point function since $P_{\zeta\gamma}=0$. 

On the other hand, $1-\sum_a r_a^2$ is proportional to the determinant of the total 2-point correlation matrix $\det P_{\varphi\varphi'}$, where $\varphi$ indicates all local fields and operators including $\zeta$. The positivity of this determinant together with \eqref{tildfNL} then implies
\be
\left(\frac65 f_{NL}\right)^2 \leq \tau_{NL}.
\ee

\centerline{*          *            *}

Let us close this section by two general remarks. First, the underlying assumption for the expansion \eqref{xi} was that the short modes are in vacuum until the long mode has crossed the horizon and become classical. This assumption can be broken for instance if the short wavelength modes get excited inside the horizon. An explicit example of this sort is axion models of inflation \cite{McAllister,Flauger} where there is another physical frequency $\omega>H$ in the problem at which the modes get excited. This case has been studied in appendix \ref{resonant}. As seen after sufficient squeezing (here $q/k \ll H/\omega$) our formalism will be applicable again. In particular, in the absence of other degrees of freedom the single-field consistency conditions hold for sufficiently squeezed correlators \cite{Flauger_squeezed}.

The second remark is regarding the resemblance between the expansion \eqref{xi} for classical long wavelength modes and the bias expansion for the distribution of halos and galaxies. In the latter case the dynamics of the short scale modes is influenced by the local observables of the long wavelength perturbations of gravitational potential (made of $\d_i\d_j\phi$) along the trajectory of the short modes. However unlike a super-horizon $\sigma$ field whose time-evolution is local and fully determined in terms of the field $\hat\sigma$ and its spatial derivatives at a single time-slice, the sub-horizon density fluctuations have long range gravitational interactions. Therefore the bias expansion won't be local if written just in terms of the initial field and its spatial derivatives (see e.g. \cite{bias}).

\section{Heavy fields $m/H>3/2$}

In this case the two solutions of the free-field equation decay with the same rate and oscillate, scaling as $\eta^{3/2\pm i\mu}$ where $\mu = \sqrt{m^2/H^2-9/4}$. Hence, the fields are not classical and we cannot expand as in \eqref{xi}. However, there is still a sense in which the correlation functions arising from the exchange of heavy fields factorize in the squeezed limit. This is because in this limit the dominant contribution arises from the production of on-shell heavy particles, whose amplitude is suppressed by a Boltzmann factor $\exp(-\pi\mu)$ for each pair of particles. In the weakly coupled case the other contributions in which no on-shell particle is created are captured by a set of effective derivative self-couplings of $\zeta$ arising from integrating out the heavy particle. These self couplings are not suppressed by any Boltzmann factor, but they lead to equilateral type non-Gaussianities. For instance, the squeezed limit bispectrum scales as $q^2$, which decays by a factor of $(q/k)^{1/2}$ faster compared to the former contribution from production of on-shell particles which goes as $(q/k)^{3/2}$. For higher statistics these on-shell particles give a sub-dominant piece but they still have a characteristically different functional form. This problem has been thoroughly studied in \cite{Noumi,Arkani-Hamed} and we will just touch upon the salient features of the squeezed limit correlators.\footnote{For a study of the impact of heavy fields on correlators involving tensor modes see \cite{Dimastrogiovanni:2015pla}.}

Instead of correlation functions in a classical background, here we should talk about the wavefunction of universe to keep track of the phase of the $\sigma$ field. Since the concept of particle is well-defined for late-time heavy fields, it is more natural to expand the wavefunction in terms of eigenstates of particle number operator as in the scattering theory. Let us choose an intermediate time $\tilde\eta$ satisfying $k^{-1}\ll -\tilde\eta\ll q^{-1}$, so that the super-horizon field $\sigma_\q$ is well-defined but the hard $k$ modes are in vacuum. The wavefunction at $\eta_0$, when all correlations are formed, can be written as 
\be\label{Psi}
|\Psi(\eta_0)\rangle = \sum_N\int \prod_{n=1}^N \frac{d^3\q_n}{(2\pi)^3}
\int D\zeta \ U(\eta_0,\tilde\eta) \ |\{\zeta\},N\sigma \rangle \ \Psi[\{\zeta\},N\sigma;\tilde\eta],
\ee
where $U(\eta_0,\tilde\eta)$ is the time-evolution operator and the sum is only over $N$-particle states of $\sigma$ with super-horizon wavelengths. Other possible degrees of freedom have been omitted. The wavefunction $|\Psi(\tilde\eta)\rangle$ depends on the sub-horizon dynamics, and if $\sigma$ exists there as a weakly coupled degree of freedom can be perturbatively calculated. However, on general grounds we expect that
\be
\Psi[\{\zeta\},N\sigma;\tilde\eta]\propto e^{-N\pi\mu/2},
\ee
since we assume that the theory is in the adiabatic vacuum deep inside the horizon, and hence particle production is a result of evolution in the expanding background. Secondly, all modes that are still deeply sub-horizon must be in vacuum:
\be
\Psi[\{\zeta\},N\sigma;\tilde\eta]\propto \delta(\zeta_\k), \qquad \text{for all}\qquad -k\tilde\eta \gg 1.
\ee
(The $i\ep$ prescription evolves such a wavefunction into the true vacuum.) On the other hand, the soft $\zeta_q$ modes are constant up to corrections of order $(q\tilde \eta)^2$, therefore
\be
\langle \zeta_\q|U(\eta_0,\tilde\eta)|\tilde\zeta_{\q}\rangle \propto \delta[\zeta_\q -\tilde \zeta_\q
+\O(q^2\tilde\eta^2)].
\ee
The correlation functions are obtained by taking averages using the wavefunction \eqref{Psi}. Keeping the minimum number of Boltzmann factors, we get:
\be
\expect{\zeta_\q\zeta_{\k_1}\zeta_{\k_2}}_{\sigma}\simeq
\expect{\zeta_{\k_1}\zeta_{\k_2}|1\sigma_\q}
\int d\zeta_\q \ \zeta_\q \ \Psi^\dagger[\zeta_\q,0\sigma;\tilde\eta] \
\Psi[\zeta_\q,1\sigma_{-\q};\tilde\eta]+\text{c.c.}
\ee
The second factor on the r.h.s. depends on the sub-horizon mixing of $\zeta_\q$ and $\sigma_\q$. We expect from the perturbative calculation that this correlation should form at $-q \eta \sim \mu\gg 1$ and to introduce another factor of $\exp(-\mu\pi/2)$. The first factor on the r.h.s. corresponds to a massive particle decaying into a pair of $\zeta_\k$ modes when their total energy red-shifts to values around the mass, i.e. $\eta\sim -\mu/k$. The super-horizon wavefunction $\sigma_q \propto (q\eta)^{3/2+i\mu}/q^{3/2}+\text{c.c.}$ then leads to an oscillating piece $\cos(\mu\log k+\phi_q)$ and a dilution factor $(q/k)^{3/2}$. The phase $\phi_q$ is fixed if we require that $\sigma$ should be in the adiabatic vacuum at very early times. As emphasized in \cite{Arkani-Hamed} the oscillations result from interference: the expectation value is calculated between the Gaussian part of the wavefunction, and the part with production of a pair of entangled $\sigma$ particles, one of them oscillating into a $\zeta_\q$ mode, and the other decaying at a much later time into a pair of short wavelength $\zeta_{\pm \k}$ modes. Fixing the other factors by approximate scale invariance, we obtain
\be\label{massive}
\expect{\zeta_\q\zeta_{\k_1}\zeta_{\k_2}}_{\sigma}\propto 
P_\zeta(q)P_\zeta(k) \left(\frac{q}{k}\right)^{3/2}\cos(\mu\log \frac{q}{k}-\phi_0).
\ee
This is enhanced in the squeezed limit by $(k/q)^{1/2}$ compared to equilateral type shapes. Moreover, the oscillating component makes this shape distinct from other contributions. We could also consider the $\sigma$-exchange contribution to the 4-point correlator:
\be
\expect{\zeta_{\k_1}\zeta_{\k_2}\zeta_{\k_3}\zeta_{\k_4}}_{\sigma}\propto e^{2\mu\pi}
\expect{\zeta_{\k_1}\zeta_{\k_2}|1\sigma_\q}
\expect{\zeta_{\k_3}\zeta_{\k_4}|1\sigma_{-\q}}+\rm{c.c.}
\ee
Again this is an interference between the Gaussian piece, and a piece which describes the following process: (i) An entangled pair of massive particles is produced in the time-dependent background. (ii) The particles move out of causal contact. (iii) They subsequently decay into several $\zeta$ fields, when the total energy of the red-shifting modes coincides with $m$. 

Finally, if the exchanged field has spin $s>0$, the squeezed limit result would be proportional to $\ep^{i_1\cdots i_s}(\q)\hat \k_{i_1}\cdots \hat \k_{i_s}$. Only the longitudinal polarization of a massive tensor field can mix with a scalar $\zeta$, e.g. via $\dot\zeta \d_{i_1}\cdots\d_{i_s}\sigma^{i_1\cdots i_s}$. Therefore, 
\be\label{spin}
\expect{\zeta_\q\zeta_{\k_1}\zeta_{\k_2}}_{\sigma}\propto 
P_\zeta(q)P_\zeta(k) \left(\frac{q}{k}\right)^{3/2}\cos(\mu\log \frac{q}{k}-\phi_0)\P_s(\hat\q\cdot\hat\k),
\ee
where $\P_s$ is the Legendre polynomial. 

\subsection{Quantum versus classical correlation}

We have seen that the squeezed limit correlation functions have an expansion in terms of a discrete set of power-laws 
\be
\begin{split}
\expect{\zeta_\q\zeta_{\k_1}\zeta_{\k_2}}&\sim P_\zeta(q)P_\zeta(k)\sum_\Delta a_\Delta \left(\frac{q}{k}\right)^{\Delta}\;, \\
\expect{\zeta_{\k_1}\zeta_{\k_2}\zeta_{\k_3}\zeta_{\k_4}}&\sim P_\zeta(k_1)P_\zeta(k_3)\sum_\Delta b_\Delta \left(\frac{q}{k}\right)^{\Delta} \qquad \text{counter-collinear}\;,
\end{split}
\ee
with $\Delta$ integers $\geq 2$ for equilateral type non-Gaussianities, real numbers corresponding to the light ($m/H<3/2$) degrees of freedom and their derivatives and products, or pairs of complex numbers corresponding to heavy degrees of freedom. This expansion is unambiguous and non-trivial and seems to be dictated just by the isometries of a quasi-de Sitter space-time. As emphasized in \cite{Arkani-Hamed} the case of heavy field exchange is fundamentally different as it corresponds to a quantum mechanical interference effect.\footnote{It was also argued that they are distinguishable due to the non-locality of the real space correlator. However, almost all terms in the above expansion lead to large distance (non-local) correlations when Fourier transformed with respect to $q$ (see the discussion of section \ref{sec:ope}).} In fact one can see that the fundamental difference between the case of heavy intermediate fields and the other cases is the difference between quantum versus classical correlations. The outcome of two short distance measurements in two far separated laboratories are correlated with each other in all cases, but in one case they are deterministic while in the other case the outcome of one measurement influences the result of the other.

This difference has a distinct imprint in the squeezed limit correlators: When the long wavelength mode becomes classical, the $\sigma$ field works as a hidden variable and the expansion \eqref{xi} is possible. As a result the dependence on the momenta on two sides of the intermediate long wavelength mode factorizes (see for instance \eqref{z3fac0} and \eqref{z4fac}), or more generally it can be written as a sum of factorized terms weighted by the correlations among various derivatives and products of $\sigma$. Different short distance measurements have no influence on one another; they are correlated because they are both influenced by a single long mode. On the other hand, the heavy field contribution is not factorizable because of the dependence of the squeezed 3-point and 4-point functions, respectively, on $\log (q/k)$ and $\log(q^2/k_{12}k_{34})$ (where $k_{ij}=|\k_i-\k_j|$) \cite{Arkani-Hamed}.

\section{Prospects in large-scale structure surveys}

In this section we will discuss the observability of various squeezed limit behaviors, when contrasted with the contribution from the gravitational interactions during the structure formation (see \cite{Flauger_squeezed} for a related analysis in the context of single-field inflation). Consider a non-Gaussian 3-point correlator whose squeezed limit behavior is
\be\label{fnl}
\expect{\zeta_\q\zeta_\k\zeta_{\k'}}=f_{NL} P(k)P(q)\left(\frac{q}{k}\right)^\Delta(2\pi)^3 \delta^3(\q+\k+\k').
\ee
The signal to noise ratio in a 3d survey of volume $V$ can be estimated as
\be
(S/N)^2=\frac{V^3}{(2\pi)^9}\int d^3\k_1 d^3\k' d^3\q 
\frac{\expect{\zeta_\q\zeta_\k\zeta_{\k'}}\expect{\zeta_\q\zeta_\k\zeta_{\k'}}}
{\expect{\zeta_\q\zeta_\k\zeta_{\k'}\zeta_\q\zeta_\k\zeta_{\k'}}}.
\ee
Using \eqref{fnl}, approximating the denominator by the Gaussian contribution, and replacing $(2\pi)^3\delta(\bsb 0)\to V$ (appropriate for finite volume surveys), we get
\be
(S/N)^2 = f_{NL}^2\frac{V}{(2\pi)^6}\int d^3\k \int d^3\q P(q) \left(\frac{q}{k}\right)^{2\Delta},
\ee
where we are calculating the signal in the squeezed limit configurations $q\ll k$. The $\k$ integral is dominated by the highest values of $k$. This is generally the case since the number of modes grows as $k^3$. For biased tracers, $k$ really corresponds to the modes that have become nonlinear. The dependence on $k$ would then be different. However, we are interested in the dependence on $q$ as it is made small so that $\zeta_\q$ mode is linear and $P(q)\propto 1/q^3$. Limiting the inner integral to $q<q_{\rm max}$ we get a contribution which is proportional to 
\be\label{S/N}
(S/N)\propto q_{\rm max}^{\Delta}.
\ee
That is, the error with which this type of non-Gaussianity can be measured by considering modes longer than $\qmax^{-1}$ scales as $\qmax^{-\Delta}$. This suggests that unless $\Delta =0$, there is more signal in larger values of $q$. 

However, there is a guaranteed level of non-Gaussianity due to the subsequent gravitational interaction of the modes that also increases as $q$ becomes bigger and competes with primordial signal. For very large values of the hard momenta $k$, this is not accurately calculable. However, in the squeezed limit its $q$ dependence must scale as
\be\label{gr}
\expect{\zeta_\q\zeta_\k\zeta_{\k'}}_{\rm gr}\propto \O(q^2) P(q),
\ee
since the short scale dynamics can only depend on the local observables made of the long mode. As $k$ is kept fixed around the largest available values and $q$ is made smaller, this gravitational contribution decays faster than \eqref{fnl} for all $\Delta<2$. Therefore, the relevant question is how the signal in modes with $q<\qmax$, i.e. equation \eqref{S/N}, compares to the ratio of \eqref{fnl} to \eqref{gr}, which scales as $\qmax ^{\Delta-2}$. One finds that for $\Delta <1$ the primordial shape is better distinguishable in the squeezed limit, while for higher values of $\Delta$ analyzing the equilateral configurations significantly improves the chance of detectability. However, unlike the squeezed limit the equilateral primordial shapes are not expected to be very distinct from gravitational contributions, hence the theoretical uncertainty in determining the latter close to the nonlinear scale poses a challenge to increasing $k$.

\section{Conclusions}

We argued that several squeezed limit properties of non-single-field inflationary models hold more generically, and can be studied in a unified fashion. The fields in a quasi-de Sitter space-time become free at super-horizon scales and follow power-law time evolution. The light fields become classical and the leading squeezed limit behaviors can be obtained from an effective parameterization of short-distance correlations in terms of a local expansion in derivatives and powers of the background long wavelength mode. We showed how the known results, such as the squeezed limit behavior in quasi-single-field models, and the unitarity constrains on multi-field models, can be derived in a more model independent way from this formalism. Heavy fields, on the other hand, have complex wavefunction at late times and are naturally described in terms of many particle states. The squeezed limit correlation functions due to the production of these heavy particles by the time-dependent background is dominated by interference effects and hence they oscillate as a function of the ratio of $k_{\rm short}/k_{\rm long}$. All these features have not much to do with the details of sub-horizon dynamics, but they essentially depend on quasi-de Sitter evolution and the existence of (not too heavy) super-horizon degrees of freedom other than metric and inflaton fluctuations. The form can be used as a general test of a quasi-de Sitter cosmic evolution. Finally, we discussed the observability of various squeezed limit scalings of the bispectrum.

\section*{Acknowledgments}

We thank Nima Arkani-Hamed, Paolo Creminelli, Juan Maldacena, and Matias Zaldarriaga for stimulating discussions. M.M. is supported by NSF Grants No. PHY-1314311 and No. PHY-0855425. M.S. is supported by the Institute for Advanced Study.

\appendix

\section{Purturbative examples of factorization}

\subsection{Fundamental fields}\label{funda}

Suppose $\zeta$ is coupled to a light field $\sigma$ via
\be
\label{H0}
\H = \H_1+\H_2 = \dot\zeta^2\sigma + \dot\zeta \sigma.
\ee
We are interested in the leading contribution of a $\sigma$ exchange to the squeezed limit of the 3-point function $\expect{\zeta_\q\zeta_{\k_1}\zeta_{\k_2}}_\sigma$:
\be\label{z30}
\begin{split}
\expect{\zeta_\q\zeta_{\k_1}\zeta_{\k_2}}_{\sigma^2}=
-2\Re\expect{\zeta_\q(\eta_0)\zeta_{\k_1}(\eta_0)\zeta_{\k_2}(\eta_0)\int^{\eta_0}_{-\infty}\frac{d\eta_1}{\eta_1^4}
\int^{\eta_1}_{-\infty}\frac{d\eta_2}{\eta_2^4}\H_1(\eta_1)\H_2(\eta_2)}\\[10pt]
+\expect{\int^{\eta_0}_{-\infty}\frac{d\eta_1}{\eta_1^4}\H_1(\eta_1)\:
 \zeta_\q(\eta_0)\zeta_{\k_1}(\eta_0)\zeta_{\k_2}(\eta_0) 
\int^{\eta_1}_{-\infty}\frac{d\eta_2}{\eta_2^4}\H_2(\eta_2)}+ \H_1(\eta_1)\leftrightarrow \H_2(\eta_2) \;,
\end{split}
\ee
where we introduced an infrared regulator $\eta_0$ since the field $\sigma$ is generically time-dependent at super-horizon scales. We will often drop the time argument of $\zeta$ if it is $\eta_0$. The main contribution to the squeezed limit is when the mixing $\H_2$ between the soft $\zeta$ and $\sigma$ fields occurs at a time around $\eta_q \sim -1/q$ much earlier than the $\H_1$ interaction at $\eta_k\sim -1/k$. Substituting \eqref{H0} in \eqref{z30}, expanding in terms of mode functions and using this approximation, we obtain
\be\label{z320}
\begin{split}
&- \zeta_\q(\eta_0)\zeta_{\k_1}(\eta_0)\zeta_{\k_2}(\eta_0) 
\int^{\eta_0}_{-\infty}\frac{d\eta_1}{\eta_1^4}
\int^{\eta_1}_{-\infty}\frac{d\eta_2}{\eta_2^4}
\zeta^*_{\k_1}(\eta_1)\zeta^*_{\k_2}(\eta_1)\sigma_q(\eta_1)
\sigma^*_q(\eta_2)\zeta^*_\q(\eta_2)+\rm{c.c.}\\[10pt]
&+ \zeta_\q(\eta_0)\zeta^*_{\k_1}(\eta_0)\zeta^*_{\k_2}(\eta_0) 
\int^{\eta_0}_{-\infty}\frac{d\eta_1}{\eta_1^4} \zeta_{\k_1}(\eta_1)\zeta_{\k_2}(\eta_1)\sigma_q(\eta_1)
\int^{\eta_0}_{-\infty}\frac{d\eta_2}{\eta_2^4}\sigma^*_q(\eta_2)\zeta^*_\q(\eta_2)+\rm{c.c.}
\end{split}
\ee
We next argue that this expression factorizes into
\be\label{z3fac}
\expect{\zeta_\q\zeta_{\k_1}\zeta_{\k_2}}_{\sigma}\simeq 
\frac{\expect{\zeta_\q\sigma_{-\q}(\eta_0)}\expect{\zeta_{\k_1}\zeta_{\k_2}\sigma_{\q}(\eta_0)}}
{P_{\sigma}(q,\eta_0)}\propto P(q)P(k)\left(\frac{q}{k}\right)^{\Delta_\sigma},
\ee
where $\Delta_\sigma=\frac{3}{2}-\sqrt{\frac{9}{4}-\frac{m^2}{H^2}}$ is the scaling of the growing mode and $P_\sigma(q,\eta_0)=|\sigma_q(\eta_0)|^2$. The two mixed correlators in the numerator are given by
\be\label{zs}
\expect{\zeta_\q\sigma_{-\q}(\eta_0)}=-i \zeta_\q(\eta_0)\sigma_{-\q}(\eta_0)
\int^{\eta_0}_{-\infty}\frac{d\eta}{\eta^4}\sigma^*_{-\q}(\eta)\zeta^*_\q(\eta)+\rm{c.c.}
\ee
and
\be\label{zzs}
\expect{\zeta_{\k_1}\zeta_{\k_2}\sigma_{\q}(\eta_0)}=-i\zeta_{\k_1}(\eta_0)\zeta_{\k_2}(\eta_0)\sigma_{\q}(\eta_0) 
\int^{\eta_0}_{-\infty}\frac{d\eta}{\eta_1^4} \zeta^*_{\k_1}(\eta_1)\zeta^*_{\k_2}(\eta_1)\sigma^*_\q(\eta_1)+\rm{c.c.}
\ee
The relation \eqref{z3fac} is then verified by noticing that: (a) The integral in \eqref{zs} is independent of $\eta_0$ as long as $-q\eta_0\ll 1$. Therefore, since the $\eta_1$ integral on the first line of \eqref{z320} is dominated by $-\eta_1\sim 1/k \ll 1/q$, we can replace the upper-bound of the $\eta_2$ integral with $\eta_0$. (b) At late times the time-dependence of the mode functions factorizes and we can replace:
\be
\sigma_q(\eta_2)\to \sigma_q^*(\eta_2)\frac{\sigma_q(\eta_0)}{\sigma_q^*(\eta_0)}.
\ee

\subsubsection{Light fields}\label{light}

For fields of dimension $\Delta_\sigma\leq 1$, the integral in \eqref{zs} is naively dominated by late times, and the approximation (a) seems to be invalid. However, this late-time dominance is spurious since it is pure imaginary and cancels in the final result. This independence from the late-time cutoff $\eta_0$ (apart from the trivial scaling in $\sigma_q(\eta_0)$) becomes manifest by reorganizing the perturbation theory as time-evolution problem \cite{Musso}. In this language there are two contributions to $\expect{\zeta_\q\sigma_{-\q}(\eta_0)}$:
\be\label{zsig}
\expect{\zeta_\q\sigma_{-\q}(\eta_0)}=\sigma^*_{-\q}(\eta_0)
\int^{\eta_0}_{-\infty}\frac{d\eta}{\eta^4} \dot G^R_{\zeta_\q}(\eta_0,\eta)\sigma_{-\q}(\eta)
+\zeta_\q(\eta_0)
\int^{\eta_0}_{-\infty}\frac{d\eta}{\eta^4} G^R_{\sigma_{-\q}}(\eta_0,\eta)\dot\zeta^*_\q(\eta),
\ee
where 
\be
G_{f_\q}^R(\eta_0,\eta)=\frac{i}{2}(f_\q(\eta_0)f^*_{\q}(\eta)-\rm{c.c.})
\ee
is the retarded propagator of the indicated field. At late times it decays as $\eta^3$, while $\dot\zeta_\q(\eta) \propto \eta^2$ and $\sigma_\q(\eta)\propto \eta^{\Delta_\sigma}$. Hence, for all $\Delta_\sigma>0$ each term in \eqref{zsig} is manifestly IR-safe, i.e. it does not depend on the upper limit of the integral as long as $-q\eta_0\ll 1$, but only the sum of the two terms is real. The case of an unprotected massless field $\Delta_\sigma =0$ is special. Here, the dependence on $\eta_0$ is real and corresponds to the possibility of the generation of a mass for $\sigma$ via its mixing with $\zeta$. 

In this formalism, the leading contribution of $\sigma$ exchange to the squeezed limit comes from two diagrams
\be
\begin{split}
\expect{\zeta_\q\zeta_{\k_1}\zeta_{\k_2}}=
\int^{\eta_0}_{-\infty}\frac{d\eta_2}{\eta_2^4}
\dot G^R_{\zeta_\q}(\eta_0,\eta_2)\sigma_{-\q}(\eta_2)
\int^{\eta_0}_{-\infty}\frac{d\eta_1}{\eta_1^4}\sigma^*_{\q}(\eta_1)\cdots\\[10pt]
+\zeta_\q(\eta_0)
\int^{\eta_0}_{-\infty}\frac{d\eta_1}{\eta_1^4}\int^{\eta_0}_{-\infty}\frac{d\eta_2}{\eta_2^4}
G^R_{\sigma_{-\q}}(\eta_1,\eta_2)\dot\zeta^*_\q(\eta_2)\cdots,
\end{split}
\ee
where dots represent the part of the diagrams corresponding to the interaction among the high momentum modes at $\eta_1\sim -1/k$. The approximation (b) can now be rephrased as
\be
G^R_{\sigma_\q}(\eta_1,\eta_2)\to G^R_{\sigma_\q}(\eta_0,\eta_2)\frac{\sigma_q(\eta_1)}{\sigma_q(\eta_0)},
\quad\text{for}\quad -q\eta_1\ll 1,
\ee
and leads to the same factorization formula \eqref{z3fac}. 

\subsection{Local operator exchange}\label{operator}

Suppose $\zeta$ is coupled to a light field $\sigma$ of mass $m<3H/2$ via
\be
\label{H}
\H = \H_1+\H_2 = \dot\zeta^2\sigma^2 + \dot\zeta \sigma^2.
\ee
We are interested in the leading contribution of a $\sigma$ loop to the squeezed limit of the 3-point function $\expect{\zeta_\q\zeta_{\k_1}\zeta_{\k_2}}_\sigma$:
\be\label{z3}
\begin{split}
\expect{\zeta_\q\zeta_{\k_1}\zeta_{\k_2}}_{\sigma^2}=
-2\Re\expect{\zeta_\q(\eta_0)\zeta_{\k_1}(\eta_0)\zeta_{\k_2}(\eta_0)\int^{\eta_0}_{-\infty}\frac{d\eta_1}{\eta_1^4}
\int^{\eta_1}_{-\infty}\frac{d\eta_2}{\eta_2^4}\H_1(\eta_1)\H_2(\eta_2)}\\[10pt]
+\Re\expect{\int^{\eta_0}_{-\infty}\frac{d\eta_1}{\eta_1^4}\H_1(\eta_1)\:
 \zeta_\q(\eta_0)\zeta_{\k_1}(\eta_0)\zeta_{\k_2}(\eta_0)
\int^{\eta_1}_{-\infty}\frac{d\eta_2}{\eta_2^4}\H_2(\eta_2)}+ \H_1(\eta_1)\leftrightarrow \H_2(\eta_2) \;,
\end{split}
\ee
where we introduced an infrared regulator $\eta_0$ since the field $\sigma$ is generically time-dependent at super-horizon scales. We will often drop the time argument of $\zeta$ if it is $\eta_0$. The contribution to the squeezed limit is insignificant unless the momenta of the virtual pair $\q_1$ and $\q_2$ are of the same order as the soft momentum $\q$, when an $\H_2$ interaction among the three soft modes occur around $\eta_q \sim -1/q$ much earlier than the $\H_1$ interaction at $\eta_k\sim -1/k$. Substituting \eqref{H} in \eqref{z3}, expanding in terms of mode functions and using this approximation, we obtain
\be\label{z32}
\begin{split}
-2\Re\int^{\eta_0}_{-\infty}\frac{d\eta_1}{\eta_1^4}
\int^{\eta_0}_{-\infty}\frac{d\eta_2}{\eta_2^4}
\Big[&\zeta_\q(\eta_0)\zeta_{\k_1}(\eta_0)\zeta_{\k_2}(\eta_0)
\zeta^*_{\k_1}(\eta_1)\zeta^*_{\k_2}(\eta_1)\zeta^*_\q(\eta_2)\\[10pt]
-\zeta_\q(\eta_0)\zeta^*_{\k_1}(\eta_0)&\zeta^*_{\k_2}(\eta_0)
\zeta_{\k_1}(\eta_1)\zeta_{\k_2}(\eta_1)\zeta^*_\q(\eta_2)\Big]\expect{\sigma^2_{\q}(\eta_1)\sigma^2_{-\q}(\eta_2)},
\end{split}
\ee
where
\be\label{sig2}
\expect{\sigma^2_\q(\eta_1)\sigma^2_{-\q}(\eta_2)}=\int_{\q_1+\q_2 =\q}
\sigma_{\q_1}(\eta_1)\sigma_{\q_2}(\eta_1)\sigma^*_{\q_1}(\eta_2)\sigma^2_{\q_2}(\eta_2).
\ee
Using the late time behavior:
\be\label{late}
\frac{\sigma(\eta_1)}{\sigma(\eta_0)}\simeq\left(\frac{\eta_1}{\eta_0}\right)^{\Delta_\sigma}\qquad 
\Delta_\sigma=\frac{3}{2}-\sqrt{\frac{9}{4}-\frac{m^2}{H^2}}
\ee
in \eqref{sig2} and substituting in \eqref{z32}, it is easy to see that it factorizes:
\be
\expect{\zeta_\q\zeta_{\k_1}\zeta_{\k_2}}_\sigma\simeq\int_{\q_1+\q_2 =\q}
\frac{\expect{\zeta_\q\sigma_{-\q_1}(\eta_0)\sigma_{-\q_2}(\eta_0)}
\expect{\zeta_{\k_1}\zeta_{\k_2}\sigma_{\q_1}(\eta_0)\sigma_{\q_2}(\eta_0)}}
{P_\sigma(q_1,\eta_0)P_\sigma(q_2,\eta_0)}.
\ee
However, one can show that the above squeezed limit 3-point function fully factorizes. Note first that because of the late time behavior \eqref{late} 
\be\label{zzss}
\expect{\zeta_{\k_1}\zeta_{\k_2}\sigma_{\q_1}(\eta_0)\sigma_{\q_2}(\eta_0)}
=f(k_1,\eta_0)P_\sigma(q_1,\eta_0)P_\sigma(q_2,\eta_0) \;,
\ee
which implies
\be\label{z33}
\expect{\zeta_\q\zeta_{\k_1}\zeta_{\k_2}}_{\sigma^2}\simeq f(k_1,\eta_0)
\expect{\zeta_\q\sigma^2_{-\q}(\eta_0)}.
\ee
Next, integrate \eqref{zzss} over $\q_1$ and $\q_2$, with $\q_1+\q_2=\q$, to obtain
\be\label{zzss2}
\expect{\zeta_{\k_1}\zeta_{\k_2}\sigma^2_{\q}(\eta_0)}
=f(k_1,\eta_0)P_{\sigma^2}(q,\eta_0).
\ee
Combining \eqref{z33} and \eqref{zzss2} we finally obtain
\be
\expect{\zeta_\q\zeta_{\k_1}\zeta_{\k_2}}_{\sigma^2}\simeq 
\frac{\expect{\zeta_\q\sigma^2_{-\q}(\eta_0)}\expect{\zeta_{\k_1}\zeta_{\k_2}\sigma^2_{\q}(\eta_0)}}
{P_{\sigma^2}(q,\eta_0)}\propto P(q)P(k)\left(\frac{q}{k}\right)^{2\Delta_\sigma}.
\ee
Let us make a few remarks:

\begin{itemize}

\item The above derivation holds only if $\Delta_\sigma\leq 1$. Otherwise, the $\eta_1$ integral in \eqref{z3} is dominated by $\eta_1\sim \eta_2$ and not by $\eta_1\sim -1/ q$: the UV divergence in the $\sigma$ loop wins over slow variation of $\zeta_q(\eta_1)$ at super-horizon scales. Nevertheless, by renormalizing the composite operator one expects to obtain the same factorized contribution.

\item Although the above formula was derived for the operator $\sigma^2$, it is expected to hold more generally for any intermediate operator $O$
\be
\expect{\zeta_\q\zeta_{\k_1}\zeta_{\k_2}}_{O}\simeq 
\frac{\expect{\zeta_\q O_{-\q}(\eta_0)}\expect{\zeta_{\k_1}\zeta_{\k_2} O_{\q}(\eta_0)}}
{\expect{O_\q(\eta_0)O_{-\q}(\eta_0)}}\propto P(q)P(k)\left(\frac{q}{k}\right)^{\Delta_O},
\ee
and the last scaling to be valid even for large mixing.

\end{itemize}

\subsection{Soft internal lines}\label{internal}

The approximation (b) of section \ref{funda} can be used to also show that the contribution of soft internal lines (or operators in general) to the correlation functions factorizes. For instance, a 4-point function in the counter-collinear limit $\k_1+\k_2=\q\to 0$ can be written as
\be\label{cc}
\expect{\zeta_{\k_1}\zeta_{\k_2}\zeta_{\k_3}\zeta_{\k_4}}_{O}\simeq 
\frac{\expect{\zeta_{\k_1}\zeta_{\k_2} O_{\q}(\eta_0)}\expect{\zeta_{\k_3}\zeta_{\k_4} O_{-\q}(\eta_0)}}
{\expect{O_\q(\eta_0)O_{-\q}(\eta_0)}}.
\ee
The intermediate operator can now be a soft $\zeta_q$ mode.

\subsection{Resonant non-Gaussianity}\label{resonant}

The above predictions always require sufficient amount of squeezing $q/k\ll 1$. This is needed to ensure that the short modes are in vacuum before the long mode crosses the horizon. An illustrative example is resonant non-Gaussianity where the factorization on soft internal lines, e.g. \eqref{cc}, can fail due to insufficient squeezing. Here the modes get excited at physical frequencies of order the resonant frequency $\omega\gg H$ and the interactions start to be important when the short modes cross this threshold. If $q/k > H/\omega$ the soft intermediate mode is not yet super-horizon when the interactions of the short modes become important. Thus, the approximation (b) cannot be made. More explicitly, the leading contribution to a 4-point function in the counter-collinear limit contains a term
\be
\begin{split}
-\zeta_\q(\eta_0)&\zeta_{\k_1}(\eta_0)\zeta_{\k_2}(\eta_0)
\int^{\eta_0}_{-\infty}\frac{d\eta_1}{\eta_1^4} V^{(3)}(\eta_1)
\int^{\eta_1}_{-\infty}\frac{d\eta_2}{\eta_2^4} V^{(3)}(\eta_2)\\[10pt]
&\zeta^*_{\k_1}(\eta_1)\zeta^*_{\k_2}(\eta_1)\zeta_\q(\eta_1)
\zeta^*_{\k_3}(\eta_2)\zeta^*_{\k_4}(\eta_2)\zeta^*_\q(\eta_2)
+\{\k_1,\k_2\}\leftrightarrow\{\k_3,\k_4\}+\rm{c.c.}
\end{split}
\ee
where $V^{(n)}(\eta)\propto \cos \omega t$ is the leading cubic vertex. The integrals are dominated when $-k\eta\sim \omega/H$ around the saddle point. Suppose $q\ll k$ but $q/k\gg H/\omega$ then we can neglect the factor $\exp(-iq\eta)$ in $\zeta_q(\eta)$ and replace
\be
\zeta_q(\eta)=\zeta_q(\eta_0) (1+iq\eta)e^{-iq\eta}\to iq\eta \zeta_q(\eta_0),
\ee
which gives an opposite contribution compared to $\zeta^*_q(\eta)\zeta_q(\eta_0)/\zeta_q^*(\eta_0)$ in the same approximation. Hence the approximation (b) does not hold in this regime.



\begin{thebibliography}{99}

\bibitem{Maldacena}
  J.~M.~Maldacena,
  ``Non-Gaussian features of primordial fluctuations in single field inflationary models,''
  JHEP {\bf 0305} (2003) 013
  [arXiv:astro-ph/0210603].

\bibitem{Creminelli:2004yq} 
  P.~Creminelli and M.~Zaldarriaga,
  ``Single field consistency relation for the 3-point function,''
  JCAP {\bf 0410}, 006 (2004)
  [astro-ph/0407059].


\bibitem{Creminelli_sct}
  P.~Creminelli, J.~Norena and M.~Simonovic,
  ``Conformal consistency relations for single-field inflation,''                
  JCAP {\bf 1207}, 052 (2012)
  [arXiv:1203.4595 [hep-th]].


\bibitem{Hinterbichler} 
  K.~Hinterbichler, L.~Hui and J.~Khoury,
  ``An Infinite Set of Ward Identities for Adiabatic Modes in Cosmology,''
  JCAP {\bf 1401}, 039 (2014)
  [arXiv:1304.5527 [hep-th]].


\bibitem{Chen} 
  X.~Chen and Y.~Wang,
  ``Large non-Gaussianities with Intermediate Shapes from Quasi-Single Field Inflation,''
  Phys.\ Rev.\ D {\bf 81}, 063511 (2010)
  [arXiv:0909.0496 [astro-ph.CO]].

\bibitem{Baumann} 
  D.~Baumann and D.~Green,
  ``Signatures of Supersymmetry from the Early Universe,''
  Phys.\ Rev.\ D {\bf 85}, 103520 (2012)
  [arXiv:1109.0292 [hep-th]].

\bibitem{Cheung:2007st} 
  C.~Cheung, P.~Creminelli, A.~L.~Fitzpatrick, J.~Kaplan and L.~Senatore,
  ``The Effective Field Theory of Inflation,''
  JHEP {\bf 0803}, 014 (2008)
  [arXiv:0709.0293 [hep-th]].

\bibitem{Baumann_equil} 
  D.~Baumann and D.~Green,
  ``Equilateral Non-Gaussianity and New Physics on the Horizon,''
  JCAP {\bf 1109}, 014 (2011)
  [arXiv:1102.5343 [hep-th]].

\bibitem{Senatore_multi} 
  L.~Senatore and M.~Zaldarriaga,
  ``The Effective Field Theory of Multifield Inflation,''
  JHEP {\bf 1204}, 024 (2012)
  [arXiv:1009.2093 [hep-th]].
  
\bibitem{Hinterbichler:2011qk} 
  K.~Hinterbichler and J.~Khoury,
  ``The Pseudo-Conformal Universe: Scale Invariance from Spontaneous Breaking of Conformal Symmetry,''
  JCAP {\bf 1204}, 023 (2012)
  [arXiv:1106.1428 [hep-th]].

\bibitem{Suyama} 
  T.~Suyama and M.~Yamaguchi,
  ``Non-Gaussianity in the modulated reheating scenario,''
  Phys.\ Rev.\ D {\bf 77}, 023505 (2008)
  [arXiv:0709.2545 [astro-ph]].

\bibitem{Smith}
  K.~M.~Smith, M.~LoVerde and M.~Zaldarriaga,
  ``A universal bound on N-point correlations from inflation,''
  Phys.\ Rev.\ Lett.\  {\bf 107}, 191301 (2011)
  [arXiv:1108.1805 [astro-ph.CO]].


\bibitem{Noumi} 
  T.~Noumi, M.~Yamaguchi and D.~Yokoyama,
  ``Effective field theory approach to quasi-single field inflation and effects of heavy fields,''
  JHEP {\bf 1306}, 051 (2013)
  [arXiv:1211.1624 [hep-th]].

\bibitem{Arkani-Hamed} 
  N.~Arkani-Hamed and J.~Maldacena,
  ``Cosmological Collider Physics,''
  arXiv:1503.08043 [hep-th].


\bibitem{Dalal} 
  N.~Dalal, O.~Dore, D.~Huterer and A.~Shirokov,
  ``The imprints of primordial non-gaussianities on large-scale structure: scale dependent bias and abundance of virialized objects,''
  Phys.\ Rev.\ D {\bf 77}, 123514 (2008)
  [arXiv:0710.4560 [astro-ph]].

\bibitem{IR} 
  L.~Senatore and M.~Zaldarriaga,
  ``On Loops in Inflation II: IR Effects in Single Clock Inflation,''
  JHEP {\bf 1301}, 109 (2013)
  [arXiv:1203.6354 [hep-th]].

\bibitem{cft} 
  D.~Green, M.~Lewandowski, L.~Senatore, E.~Silverstein and M.~Zaldarriaga,
  ``Anomalous Dimensions and Non-Gaussianity,''
  JHEP {\bf 1310}, 171 (2013)
  [arXiv:1301.2630].

\bibitem{LopezNacir:2011kk} 
  D.~Lopez Nacir, R.~A.~Porto, L.~Senatore and M.~Zaldarriaga,
  ``Dissipative effects in the Effective Field Theory of Inflation,''
  JHEP {\bf 1201}, 075 (2012)
  [arXiv:1109.4192 [hep-th]].

\bibitem{Nacir} 
  D.~Lopez Nacir, R.~A.~Porto and M.~Zaldarriaga,
  ``The consistency condition for the three-point function in dissipative single-clock inflation,''
  JCAP {\bf 1209}, 004 (2012)
  [arXiv:1206.7083 [hep-th]].

\bibitem{Chen:2009zp} 
  X.~Chen and Y.~Wang,
  ``Quasi-Single Field Inflation and Non-Gaussianities,''
  JCAP {\bf 1004}, 027 (2010)
  [arXiv:0911.3380 [hep-th]].


\bibitem{Assassi} 
  V.~Assassi, D.~Baumann and D.~Green,
  ``Symmetries and Loops in Inflation,''
  JHEP {\bf 1302}, 151 (2013)
  [arXiv:1210.7792 [hep-th]].
          
\bibitem{Weinberg_adia}
  S.~Weinberg,
  ``Adiabatic modes in cosmology,''    
  Phys.\ Rev.\ D {\bf 67}, 123504 (2003)
  [astro-ph/0302326].

\bibitem{anal} 
  D.~Baumann, D.~Green, H.~Lee and R.~A.~Porto,
  ``Signs of Analyticity in Single-Field Inflation,''
  arXiv:1502.07304 [hep-th].


\comment{
\bibitem{WeinbergI} 
  S.~Weinberg,
  Phys.\ Rev.\ D {\bf 74}, 023508 (2006)
  [hep-th/0605244].

\bibitem{WeinbergII} 
  S.~Weinberg,
  Phys.\ Rev.\ D {\bf 72}, 043514 (2005)
  [hep-th/0506236].
}


\bibitem{McAllister} 
  L.~McAllister, E.~Silverstein and A.~Westphal,
  ``Gravity Waves and Linear Inflation from Axion Monodromy,''
  Phys.\ Rev.\ D {\bf 82}, 046003 (2010)
  [arXiv:0808.0706 [hep-th]].

\bibitem{Flauger} 
  R.~Flauger, L.~McAllister, E.~Pajer, A.~Westphal and G.~Xu,
  ``Oscillations in the CMB from Axion Monodromy Inflation,''
  JCAP {\bf 1006}, 009 (2010)
  [arXiv:0907.2916 [hep-th]].

\bibitem{Flauger_squeezed} 
  R.~Flauger, D.~Green and R.~A.~Porto,
  ``On squeezed limits in single-field inflation.  Part I,''
  JCAP {\bf 1308}, 032 (2013)
  [arXiv:1303.1430 [hep-th]].


\bibitem{bias} 
  M.~Mirbabayi, F.~Schmidt and M.~Zaldarriaga,
  ``Biased Tracers and Time Evolution,''
  arXiv:1412.5169 [astro-ph.CO].
  
\bibitem{Dimastrogiovanni:2015pla} 
  E.~Dimastrogiovanni, M.~Fasiello and M.~Kamionkowski,
  ``Imprints of Massive Primordial Fields on Large-Scale Structure,''
  arXiv:1504.05993 [astro-ph.CO].

\bibitem{Musso} 
  M.~Musso,
  ``A new diagrammatic representation for correlation functions in the in-in formalism,''
  JHEP {\bf 1311}, 184 (2013)
  [hep-th/0611258].
\end{thebibliography}
\end{document}